\documentclass[twocolumn]{aastex631}
\usepackage[version=3]{mhchem}

\newcommand{\red}[1]{{\color{black}#1}}

\newcommand{\blue}[1]{{\color{black}#1}}

\shorttitle{SIGO formation with \ce{H2} chemistry}
\shortauthors{Nakazato et al.}

\graphicspath{{./}{figures/}}
\newcommand{\figdir}{./}
\begin{document}

\title{H$_2$ cooling and gravitational collapse of supersonically induced gas objects}
\author[0000-0002-0984-7713]{Yurina Nakazato}
\affiliation{Department of Physics, The University of Tokyo, 7-3-1 Hongo, Bunkyo, Tokyo 113-0033, Japan}

\author[0000-0001-6246-2866]{Gen Chiaki}
\affiliation{Astronomical Institute, Tohoku University, 6-3, Aramaki, Aoba-ku, Sendai, Miyagi 980-8578, Japan}

\author[0000-0001-7925-238X]{Naoki Yoshida}
\affiliation{Kavli Institute for the Physics and Mathematics of the Universe (WPI), UT Institute for Advanced Study, The University of Tokyo, Kashiwa, Chiba 277-8583, Japan}
\affiliation{Research Center for the Early Universe, School of Science, The University of Tokyo, 7-3-1 Hongo, Bunkyo, Tokyo 113-0033, Japan}

\author[0000-0002-9802-9279]{Smadar Naoz}
\affiliation{Department of Physics and Astronomy, UCLA, Los Angeles, CA 90095, USA}
\affiliation{Mani L. Bhaumik Institute for Theoretical Physics, Department of Physics and Astronomy, UCLA, Los Angeles, CA 90095, USA}
\author[0000-0002-4227-7919]{William Lake}
\affiliation{Department of Physics and Astronomy, UCLA, Los Angeles, CA 90095, USA}
\affiliation{Mani L. Bhaumik Institute for Theoretical Physics, Department of Physics and Astronomy, UCLA, Los Angeles, CA 90095, USA}
\author[0000-0003-4962-5768]{Yeou S. Chiou}
\affiliation{Department of Physics and Astronomy, UCLA, Los Angeles, CA 90095}
\affiliation{Mani L. Bhaumik Institute for Theoretical Physics, Department of Physics and Astronomy, UCLA, Los Angeles, CA 90095, USA}

\begin{abstract}
We study the formation and gravitational collapse of supersonically induced gas objects
(SIGOs) in the early universe. We run cosmological hydrodynamics simulations of SIGOs, including relative streaming motions between baryons and dark matter.
Our simulations also follow nonequilibrium
chemistry and molecular hydrogen cooling in primordial
gas clouds. A number of SIGOs are formed in the run with fast-streaming motions of 2 times the rms of the cosmological velocity fluctuations. We identify a particular gas cloud that condensates by H$_2$ cooling without being hosted by a dark matter halo. The SIGO remains outside the virial radius of its closest halo, and it becomes Jeans unstable when the central gas-particle density reaches $\sim 100~{\rm cm}^{-3}$
with a temperature of $\sim$ 200 K.
The corresponding Jeans mass is $\sim 10^5 M_{\odot}$, and thus the formation of primordial stars or a star cluster is expected in the SIGO.
\end{abstract}


\section{Introduction} \label{sec:intro}
A broad range of observations, including the measurements of the cosmic microwave background radiation anisotropies and statistics of the large-scale galaxy distribution, has established the so-called standard cosmological model, in which the energy content of the universe is dominated by dark matter (DM) and dark energy. According to the model, cosmic structure develops hierarchically,
with small subgalactic objects forming first in the early universe. It is thought that the first luminous objects were formed at $z \sim 30-20$,
when the Dark Ages ended.

The first stars were born under peculiar physical conditions. There exist relative streaming motions between the gas and dark matter, which originate from
baryon acoustic oscillations in the photon-baryon fluid \citep{Tseliakhovich_Hirata:2010}.
The typical value of the relative velocity is $v_\mathrm{bc, rec}\sim 30 ~{\rm km}{\rm s}^{-1}$ at the recombination epoch, which is about five times greater than the sound speed (hence supersonic). Also, the velocity field is coherent over a few megaparsec scales.
\red{The relative streaming velocity (SV) causes a significant impact on structure formation in the early universe. In order to examine the effect of the SV on early star formation, several hydrodynamics simulations have been performed with incorporating SV. It has been shown that SV lowers the gas fraction in low-mass DM halos and prevents or delays the formation of star-forming gas clouds \citep{Greif:2011, Tseliakhovich:2011, Fialkov:2012, Naoz:2012, Bovy_Dvorkin:2013, Schauer:2019}.  The delay of star formation due to SV brings about different physical conditions
than previously studied \citep{Tanaka_Li:2014,  Hirano:2017, Hirano:2018, Kulkarni:2020, Schauer:2020}. For example, \citet{Hirano:2017} find that SV generates strong turbulence in massive gas clouds and enhances the formation of supermassive stars.
These previous studies focused on how SV affects the formation and evolution of gas clouds that are hosted by DM halos.}
%Also, SV can affect the fluctuation of 21cm powerspectrum \citep{McQuinn:2012}.

There is an interesting possibility that SV enables baryon density peaks to form outside of DM halos \citep{Naoz_Narayan:2014}.
Such gaseous objects generated by the SV are called supersonically induced gas objects (SIGOs), which may be a new type of progenitor of primordial star clusters. \red{SIGOs have already been identified in several recent simulations.}
For example,
\citet{Popa:2016} used hydrodynamic simulations with SV to show that
gas-dominated objects
are formed in the early universe. Later, \citet{Chiou:2019, Chiou:2021} incorporated atomic hydrogen cooling in their hydrodynamics simulations and showed that a number of dense SIGOs are formed. However, it remains unclear whether or not, and how, stars are actually formed in SIGOs.
% They also proposed that SIGO could be a progenitor of globular clusters, which are known to be nearly DM-free objects(\citet{Conroy:2011}, \citet{Ibata:2013}).

  Molecular hydrogen cooling may play a vital role in the early universe. \ce{H2} cooling can lower
  the temperature of primordial gas clouds to $\sim$200\,K.
  Gas clouds with the corresponding Jeans mass of
  $\sim 1000 M_{\odot}$ become gravitationally unstable to collapse, further forming stars
  \citep{Yoshida:2008}.
In the present paper, we perform cosmological simulations with SV in order to study the formation
and evolution of SIGOs. We incorporate \ce{H2} cooling and
examine if SIGOs can cool and condense to form
stars. The rest of the paper is structured as follows. In Section \ref{sec:setup} we detail the simulation setup. In Section \ref{sec:result} we investigate how \ce{H2} chemistry affects SIGOs' formation under the gas-flowing environment and follow SIGOs' collapse. We give our concluding remarks in Section \ref{sec:discussion}.

Throughout the present paper, we adopt the standard $\Lambda$CDM cosmology with $\Omega_\Lambda = 0.73, \Omega_{\rm m} = 0.27, \Omega_{\rm b} = 0.044,$ and $h = 0.71$.

\section{Method} \label{sec:setup}
\subsection{Cosmological simulations}\label{subsec:cosmologial_simulations}
%\section{Manuscript styles} \label{sec:style}
We use the cosmological simulation code AREPO \citep{Springel:2010}.
We first run parent simulations with employing $512^3$ DM particles with a mass of $1.9\times 10^3 M_\odot$ and $512^3$ Voronoi mesh cells with a mass of 360 $M_\odot$. The simulation box size is $1.4$ comoving $h^{-1}$~Mpc on a side.
We use a modified version of the CMBFAST code \citep{Seljak:1996} to generate the transfer functions for the initial conditions. The transfer function calculations incorporate the first-order scale-dependent temperature fluctuations \citep{Naoz_Barkana:2005} and the effect of SV.
As in \citet{Chiou:2019, Chiou:2021}, we generate the initial conditions by setting a large density fluctuation amplitude of $\sigma_8 = 1.7$. This choice is aimed at simulating a rare, overdense region in a large volume where structure forms early.

We run four simulations listed in Table \ref{table:setup}. The naming convention is as follows: v2 or v0 represents with/without SV, and “H2” or “H” denotes whether \ce{H_2} cooling is turned on. For the Runs 2vH2 and 2vH, we add a coherent SV with $2 \sigma =11.8 ~{\rm km}~{\rm s}^{-1}$ in the $x$ direction to the baryonic component at
the initial redshift of $z_{\rm ini} = 200$.
We run the parent simulations to $z = 25$.

\begin{table}[htbp]
\centering
\begin{tabular}{ccc}
Run & $v_\mathrm{bc}$ & \ce{H_2} Cooling \\\hline
0vH2 & 0            & Yes        \\\hline
0vH  & 0            & No         \\\hline
2vH2 & 2$\sigma$    & Yes        \\\hline
2vH  & 2$\sigma$    & No         \\\hline

\end{tabular}
 \caption{Simulation parameters.}
  \label{table:setup}
\end{table}

\subsection{ Chemistry and cooling}\label{subsec:grackle}
We follow nonequilibrium chemical reactions and the associated radiative cooling in a primordial gas. We use the chemistry and cooling library GRACKLE \citep{grackle:2017, Chiaki:2019}.
The chemistry network includes 49 reactions for 15 primordial species: e, H, \ce{H+}, \ce{He}, \ce{He+},
He$^{++}$, H$^{-}$, H$_2$, H$_2^{+}$, D, D$^{+}$, HD, HeH$^{+}$, D$^{-}$, and HD$^{+}$.
%As \ce{H2} reactions, we include collisional ionization/recombination, $\mathrm{H}^-$/\ce{H+} processes and three-body reactions.
%We adopt the \ce{H2} cooling rate of \citet{Chiaki:2019} that is based on result of \citet{Omukai:2000}.
We include \ce{H2} and HD molecular cooling. The radiative cooling rate by \ce{H2} is calculated by
following both rotational and vibrational transitions \citep{Chiaki:2019}.

\begin{figure*}[ht!]
\centering
\includegraphics[scale = 0.69, clip]{\figdir/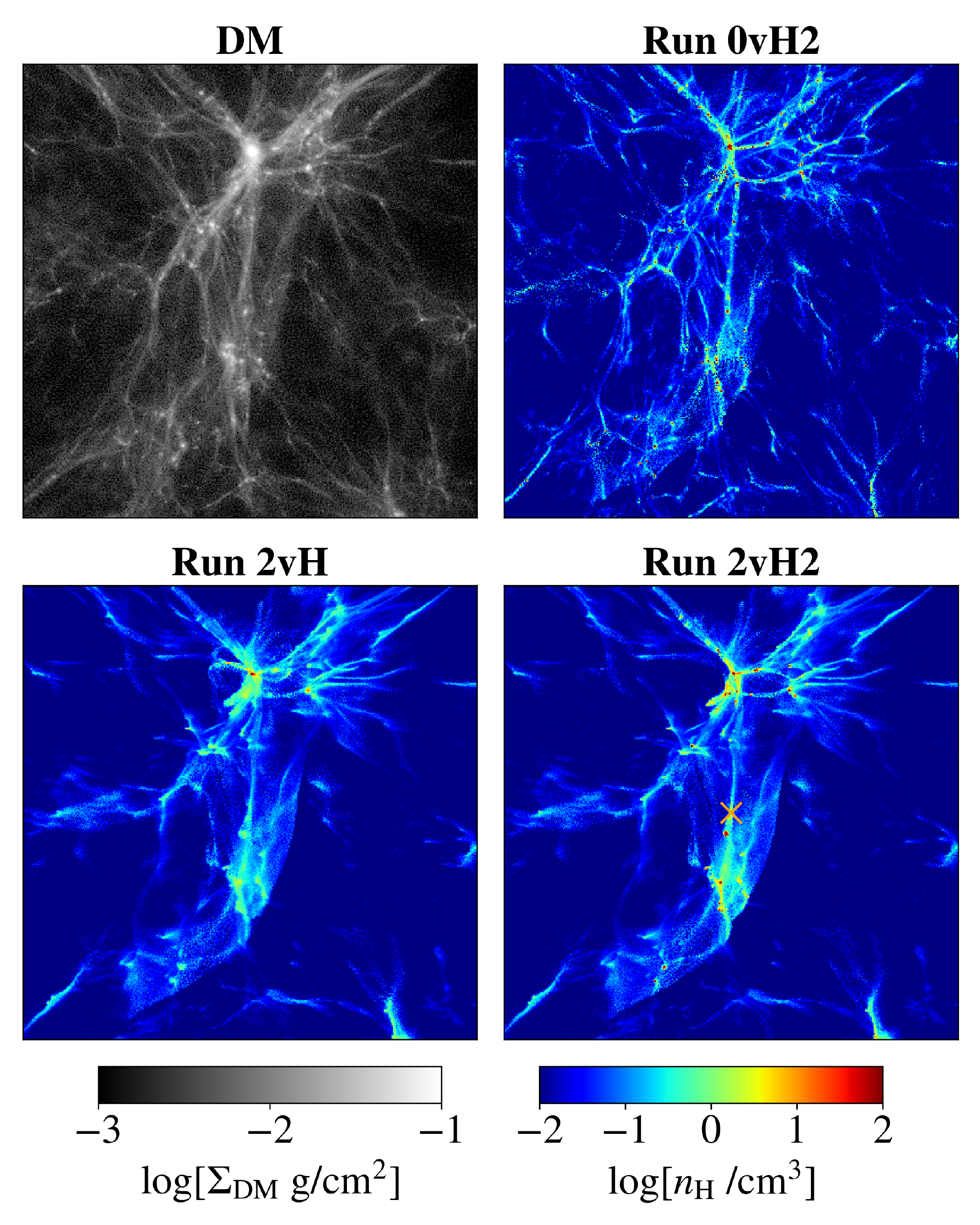}
\caption{\red{Large-scale distributions of gas and dark matter. The upper-left color map shows the DM column density in Run 0vH2.}
We plot the \red{projected} gas number density distributions in our simulations without SV (Run 0vH2, upper right) and with SV (Run 2vH2, bottom right). Run 2vH (bottom left) includes SV but not \ce{H2} cooling.
We use the outputs at $z = 25$. \red{The DM distribution is almost the same in all the three runs.}
 \red{Each color map shows a region with
a side length and a depth of 400 comoving kpc. In Run 2vH2, SIGO (S1) is located at the center of the figure, as indicated by an X mark.}
%In each set, the upper panel shows a region with a side length of 400 comoving kpc, and the bottom panel shows the enlargement of the lower right portion of the upper panel.
\label{fig:large_struct}}
\end{figure*}
%{\gc{The titles of the left and middle panels are wrong. => "0vH2" and "2vH2", respectively.}}

\subsection{Definition of SIGOs} \label{subsec:SIGO_difinition}
We first identify nonlinear objects such as DM halos
in essentially the same manner as in \citet{Popa:2016} and \citet{Chiou:2018}.
We run a friends-of-friends (FOF) group finder with a linking length of 0.2 times the mean particle separation \citep{FOF:2009}.  The smallest DM halos contain typically $\sim 300$ DM particles. We also run the FOF finder to the
gas components in order to identify "gas-only" objects that contain over 100 gas cells. \red{The minimum mass of the gas halos and the DM halos are $3.68 \times 10^4\, M_\odot$ and  $6.04 \times 10^5\, M_\odot$ respectively.}

We calculate the gas mass fraction for the identified DM halos and gaseous clouds.
Many of the detected gas clouds are filamentary, and thus it is not appropriate to measure the baryon fraction assuming spherical symmetry. We adopt an ellipsoid approximation introduced in \citet{Popa:2016}. We outline this procedure here for completion.
First, for each gas halo/cloud identified by our FOF finder, we consider an ellipsoidal surface that surrounds all of the constituent gas cells.
Then the major axis of the ellipsoid is reduced by a small amount of $0.5$\%. We repeat this procedure until the condition
\begin{equation}
   \frac{a_\mathrm{gas,n}}{a_\mathrm{gas,0}} > \frac{N_\mathrm{gas,n}}{N_\mathrm{gas, 0}} \ ,
\end{equation}
or $N_\mathrm{gas,n}/ N_\mathrm{gas, 0} < 0.8$, is met, where $a_\mathrm{gas,0}$ is the major axis of the original ellipsoid and $a_\mathrm{gas, n}$ is that of the ellipsoid after the $n$th iteration.
Similarly, $N_\mathrm{gas,0}$ and $N_\mathrm{gas, n}$ are the number of gas cells.
This iterative procedure successively shrinks the long axis of a gas halo while retaining the high-density region.
We then calculate the gas fraction of each ellipsoid as
\begin{equation}
f_\mathrm{gas} = \frac{M_\mathrm{gas}}{M_\mathrm{gas} + M_\mathrm{DM}}
\end{equation}
where $M_\mathrm{gas}$ and $M_\mathrm{DM}$ are the masses of the gas cells and DM particles within the defined ellipsoid, respectively. \blue{We note that the mass center of a SIGO is taken as the center of its ellipsoid, whereas \citet{Schauer:2021} take the highest-density point as a center of a gas clump. We have checked the both coordinates of our SIGOs, and have found that the deviation is typically a small fraction ($\sim 0.1$) of the size of the ellipsoid.}
%\red{We note that the densest part of a SIGO is often found slightly deviated from the mass center of the ellipsoid, but the deviation is typically a small fraction ($\sim 0.1$) of the ellipsoid's size.}

Finally, we identify SIGOs that satisfy the following two conditions:
(1) The mass center of the gas cells is outside the virial radii of its closest DM halo(s), and (2) there are at least 32 gas cells and the gas mass fraction is greater than 0.6 in each defined ellipsoid.
We note that the threshold value here is larger than the 0.4 adopted in \citet{Chiou:2021}.
We have found that, when the critical value is set to $0.4$, filamentary structures tend to be identified as SIGOs, especially in Run 0vH2, and many SIGOs are misidentified.
We thus set $f_\mathrm{gas, crit} = 0.6$.

\subsection{High-resolution simulation with smaller Box Size} \label{subsec:high_reso_simulation}
\red{We are not able to follow the evolution of SIGOs to $z < 25$ in our parent simulation. This is because the gravitational and hydrodynamical time scales become too short in other high-density star-forming regions in the simulated volume, and the calculations do not proceed.

We thus reconfigure and continue the simulation
by ignoring the evolution of the other halos and gas clouds far from a target SIGO, but
with increasing the mass resolution
in and around it.
In practice, we cut out a cubic region of 10 physical kpc on a side centered at the SIGO.
%In order to avoid
%following the evolution of gas clouds in the outer regions, we set the density of gas cells outside of three times the virial radius of SIGO as the critical density.
We then advance the "high-resolution" simulation by performing refinement of gas cells to ensure that the local Jeans length is always resolved with at least $64$ cells. The simulation results are shown in \ref{subsec:result_high_reso}

In both our parent and high-resolution simulations, we do not include a Lyman-Werner radiation background because the background intensity
is expected to be significant only at $z<15$ according to the cosmological simulations of \citet{Agarwal:2012}.}

\begin{figure*}[ht!]
\centering
\includegraphics[width = \linewidth, clip]{\figdir/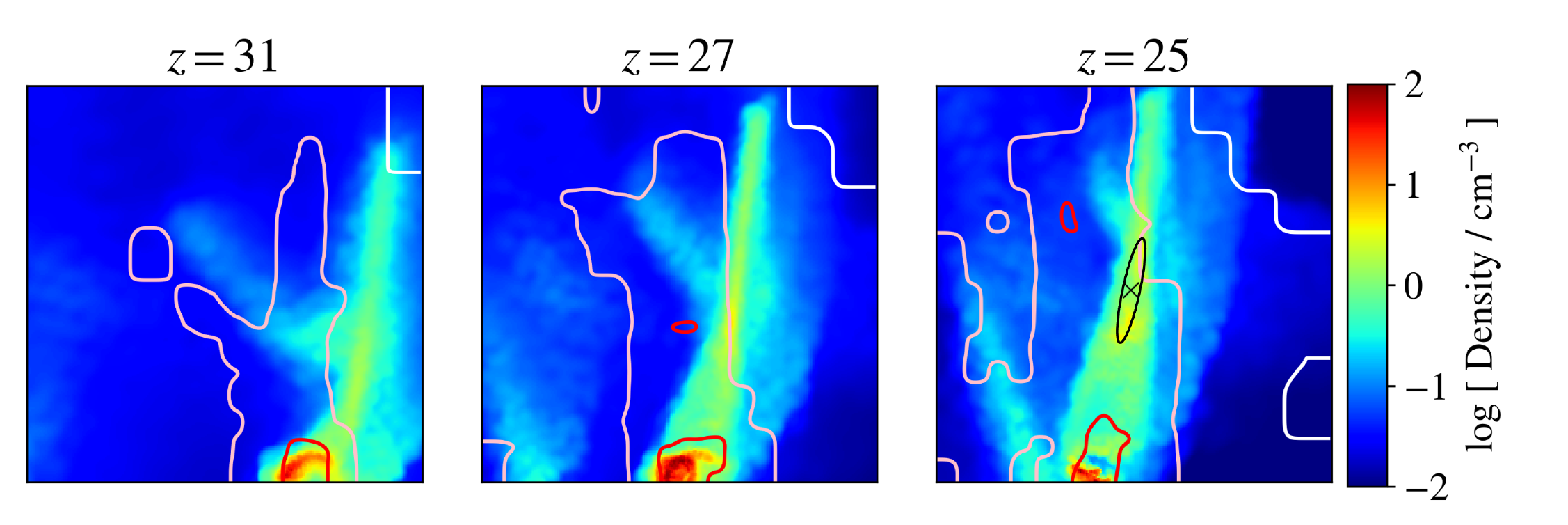}
\caption{The \red{projected} distribution of gas and dark matter around gas cloud S1 from $z = 31$ to $z = 25$. The dense gas cloud S1 is formed at $z = 25$, and appears in the center of the right panel. The initial SV is imposed in the direction from left to right.
The color map indicates the gas number density (see the color bar on the right), whereas the contours show the DM mass density distribution. The contour lines with white, pink, and red indicate 2, 20, and 200 times the critical density, respectively. \blue{The cross mark indicates the mass center of S1 and the black line indicates the S1 ellipsoid.}
We show the region with a side length of 40 comoving kpc.
\label{fig:SIGO_evo}}
\end{figure*}
\section{Result} \label{sec:result}
Figure \ref{fig:large_struct} shows the \red{projected DM distribution in Run 0vH2, and} the gas distributions in Runs 0vH2, 2vH2, and 2vH at $z = 25$. Both Runs 2vH2 and 0vH2 include \ce{H2} chemistry and cooling but Run 0vH2 does not include SV. Run 2vH includes SV but not \ce{H2} chemistry.
The effect of SV is clearly seen as coherent stream features from left to right in Runs 2vH2 and 2vH. Less small-scale structure is
seen in Run 2vH2 and 2vH than in Run 0vH2 \citep{Schauer:2020, Schauer:2021}. The comparison emphasizes the enhanced gas condensation by \ce{H2} cooling. At $z = 25$, we find the most abundant gas density peak in Run 0vH2, followed in order by 2vH2 and 2vH.
%\footnote{The number of gas density peaks is the one of gas clumps detected by the FOF algorithm. Thus, these high density regions includes both gas clumps being hosted by DM halos and SIGOs.}.
SV with 2$\sigma$ strongly suppresses gas clumping, but \ce{H2} cooling enables gas condensation, as can be seen clearly by
comparing Run 2vH2 and Run 2vH.

\red{At $z = 25$, there are 831 isolated density peaks with $n > 10^3 ~{\rm cm}^{-3}$ in Run 2vH2, and the corresponding number is 642, 2508, and 1758 in Run 2vH, 0vH2, and 0vH, respectively. There are significantly less SIGOs in each run. Without SV, only five SIGOs are found in Run 0vH2 at $z = 25$. This should be compared with 68 and 36 SIGOs in Run 2vH2 and 2vH. SV causes the gas density peaks to move fast with respect to the underlying DM, and some gas clouds start contracting while being outside of any DM halo. Note also that twice as many SIGOs are formed in Run 2vH2 than in Run 2vH at $z = 25$. This is due to enhanced gas condensation by \ce{H2} cooling.}

\subsection{Gravitational collapse of a SIGO} \label{subsec:SIGO_evo}
It is important to identify and study in detail SIGOs that can actually bear stars, if any exist.
In this section, we discuss the evolution of a particular gas cloud "S1" that is identified in Run 2vH2. S1 is a self-gravitating cloud, which eventually collapses without being hosted by a DM halo. \red{We select this cloud S1 because of its large mass and its distance from the nearby halo(s) so that it is less affected dynamically by the local structure than other SIGOs in Run 2vH2.}

Figure \ref{fig:SIGO_evo} shows the large-scale environment around S1
and its time evolution from $z = 31$ to $z = 25$.
%We project number density distribution of gas around S1 onto $x-y$ plane and show it by colormap. Also, we show mass density distribution of DM particles by contours. The contour lines with white, pink, red indicate 2, 20, 200 times of critical density.
S1 appears as a small gas clump in the center of the right panel ($z =25$). \red{When we fit S1 as an ellipsoid at $z=25$, the length along the major axis is 1.15 kpc.}

The large filamentary structure including S1 (progenitor) is formed
by $z=31$.
%\begin{equation}
%v_\mathrm{bc} \sim 2\sigma_\mathrm{bc, %rec}\frac{1+z}{1 + z_\mathrm{rec}} = 1.9 %\,\mathrm{km~s^{-1}}.
%\end{equation}
The filamentary structure is shaped by
the underlying, but slightly displaced,
DM filament shown by
the contours in Figure \ref{fig:SIGO_evo}. The gas density peak is shifted to the right from the peak of the DM owing to SV in the direction to the right. Interestingly, the gas filament starts
{\it moving back toward} the gravitational potential of the DM structure from $z=31$ to 25, in the {\it opposite} direction to the SV.
\red{The mean velocity of the filament at $z=31$
is $\sim 7$ km s$^{-1}$, which is much larger than the mean SV of $v_\mathrm{bc} = 1.9\,\mathrm{km~s^{-1}}$}.

At $z = 25$, there is no obvious DM halo around the densest part where S1 is located. \red{The distance between S1 and the closest DM halo is 1.1 physical kpc, which is over 4 times larger than the virial radius of the DM halo, which has a mass of $6.16 \times 10^6 \, M_\odot$.} At that time, the local baryon fraction of S1 is $f_\mathrm{gas}$ = 0.67.
The central gas density of S1 is 8.0 ${\rm cm}^{-3}$ and the temperature is $\sim 500$ K. With these properties, S1 is still stable against gravitational collapse. To see this more quantitatively, we calculate the ratio of the enclosed gas mass to the Jeans mass,
\begin{equation}
M_{\rm J} = \frac{\pi}{6}\frac{c^3_s}{G^{3/2}\rho^{1/2}} \ ,
\end{equation}
where $\rho$ is the density, $c_s$ is the speed of sound and $G$ is the gravitational constant. \red{ We use mass-weighted mean values $\rho~ (<r)$ and $c_s~ (<r)$ within radius $r$.}

 We find that the enclosed mass of S1 is less than the Jeans mass at all radii at $z=25$. This is consistent with the result of a recent study by \citet{Schauer:2021}, who also find similar dense gas clumps located outside of its closest DM halos. However,  \citet{Schauer:2021} argue that none of the gas clumps satisfies the Jeans instability condition for collapse because of the low gas number density of typically $\sim 1-10 \,{\rm cm}^{-3}$. As we shall show in the next section, the particular gas cloud S1 in our simulation enters runaway collapse via Jeans instability.

\begin{figure*}[ht!]
\centering
\includegraphics[width = \linewidth, clip]{\figdir/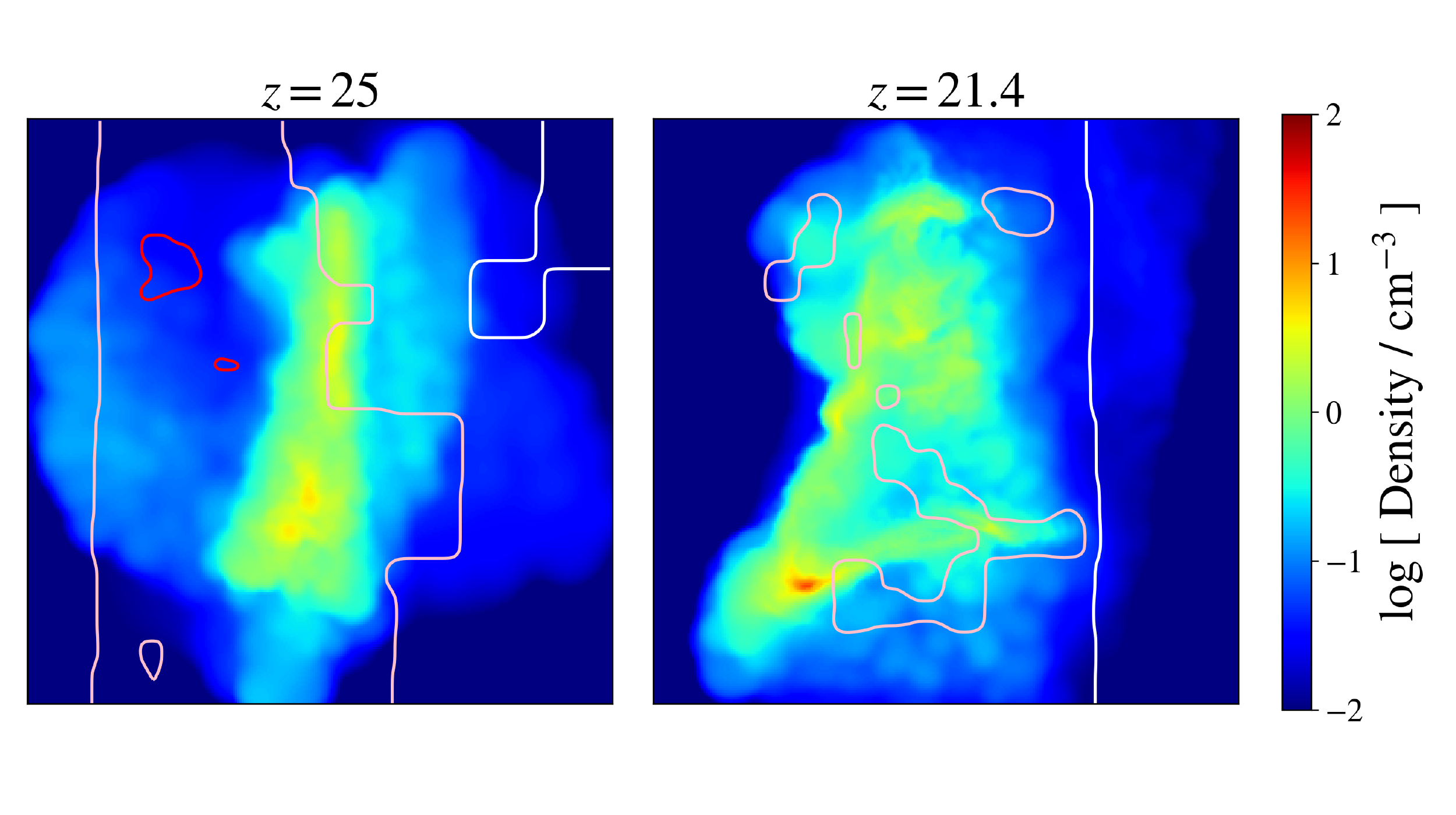}
\vspace{-1.3cm}
\caption{The \red{projected} gas distribution around S1 at $z=25$ and at $z = 21.4$, when S1 becomes Jeans unstable. The contour lines with white, pink, and red indicate the DM density of 2, 20, and 200 times the critical density. Both panels show the region around S1 with a side length of 1 physical kpc. The \blue{high-resolution simulation} shown here is performed with the refinement of gas cells.\label{fig:SIGO_evolution_zoomin} }
\end{figure*}

It is worth noting here that there is another massive and dense gas clump, appearing in the bottom center of the panels in Figure \ref{fig:SIGO_evo}. The gas clump is already forming at $z = 31$; it is located inside the virial radii of its closest DM halo at $z = 25$. Thus, it is similar to ordinary primordial gas clouds formed in DM mini halos.
%Its mass and gas fraction are $8.2 \times 10^5 \,M_\odot$ and mass ratio of gas cloud to the host DM halo is $0.21$ at  $z=25$, and thus it is similar to ordinary primordial gas clouds formed in DM mini-halos.
The DM halo is located about 30 comoving kpc away from S1 (Figure \ref{fig:SIGO_evo}; there is also a $\sim 20$ kpc separation in the $z$ direction) and is unlikely to affect the formation of S1 either via feedback effects even if stars are formed there \citep{Susa:2007} or via tidal effect, except that the whole filamentary structure surrounding S1 is slowly pulled by the gravity of the gas and DM halo.

We have checked whether a similar SIGO exists at or near the same position as S1 in Run 0vH2 and 2vH.
In Run 2vH, there is neither clump nor density peak corresponding to S1. This is likely because \ce{H} cooling is not efficient in diffuse, warm filaments.
In 0vH2, \ce{H2} cooling is effective, and there is a gas cloud similar to S1, but it is hosted by a DM halo and thus has a low baryon fraction of $f_\mathrm{gas} = 0.18$; it is not a SIGO. We conclude that S1 in our Run 2vH2 is formed through the combined effects of the SV and \ce{H2} cooling.

\begin{figure}[ht!]
\begin{center}
\;\;\;\;\;\includegraphics[scale=0.605]{\figdir/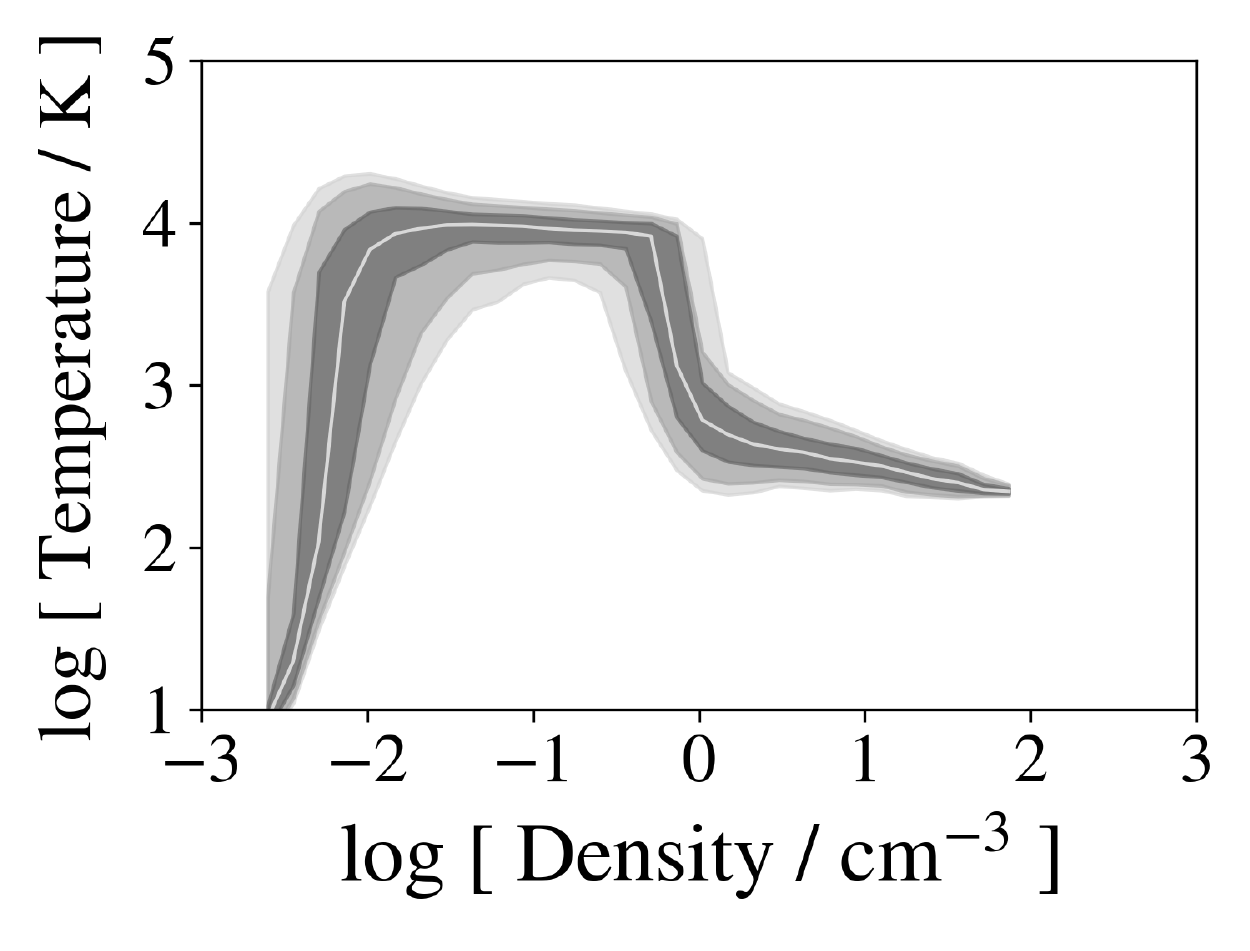}
\includegraphics[scale=0.64]{\figdir/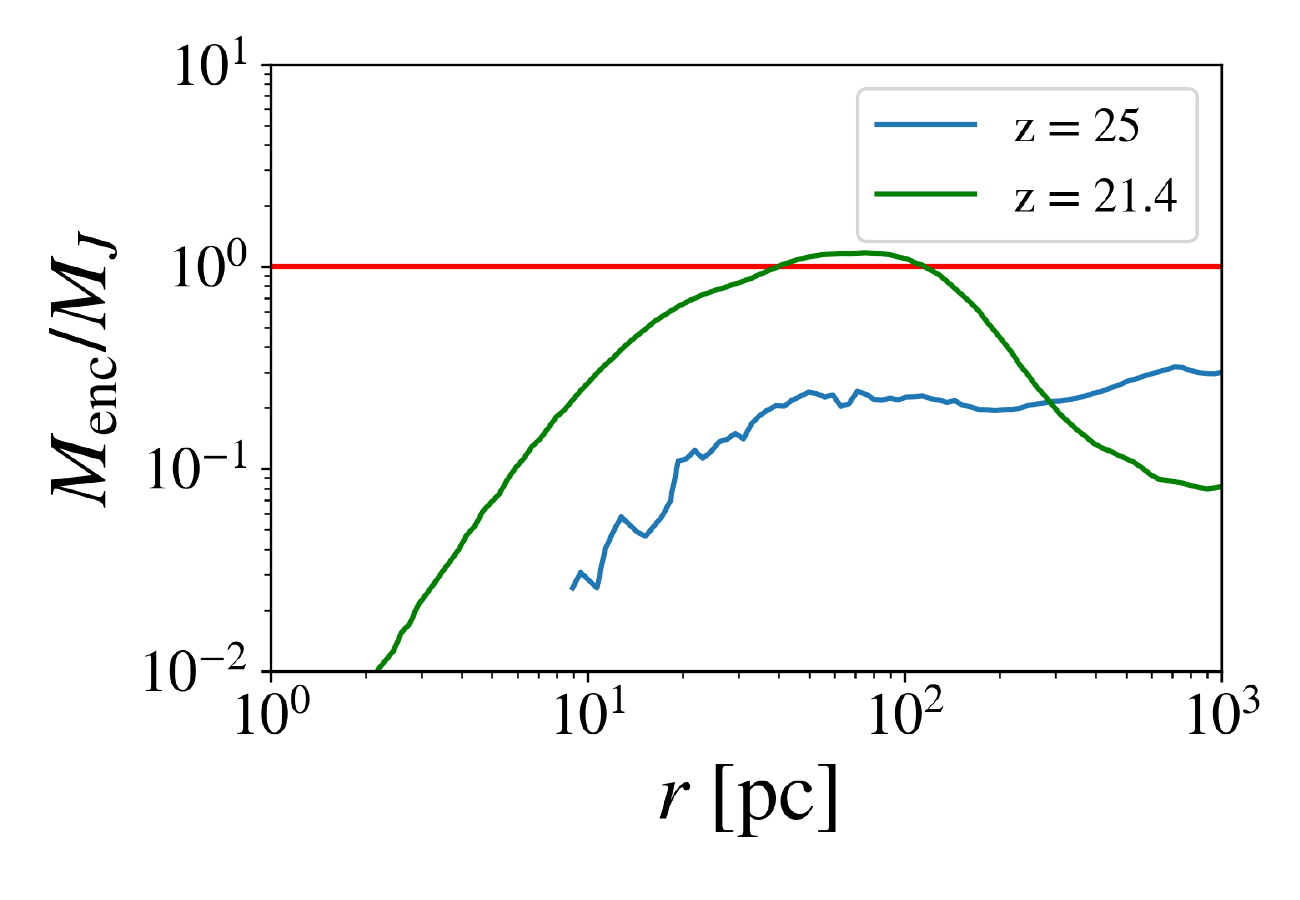}%\;\;\;\;\;\;\;\;\;\;\;\;\;\;
\end{center}
\vspace{-0.7 cm}
\caption{Top: the phase distribution of gas of S1 at $z = 21.4$. \red{The white line shows the median at each density bin. The gray bands are the qth percentiles of the gas cells with $q=5, 10, 25, 50 \mathrm{(median)}, 75, 90,$ and $95$. We plot the gas cells with $n_\mathrm{gas} > 10^{-2.6}\, \rm{cm}^{-3}$ around S1. %(\red{extracted the particles around the simulation boundaries in the t-rho plot.})
}
Bottom: radial profiles of the ratio of the enclosed gas mass to the Jeans mass. $M_\mathrm{enc}/M_J\geqq 1$ indicates that the SIGO S1 is Jeans unstable.\label{fig:zoomin_t_rho}}
\end{figure}

\subsection{High-resolution simulation with refinement}\label{subsec:result_high_reso}
\red{After we locate S1 at $z=25$ in the large-volume simulation (Section \ref{subsec:cosmologial_simulations}),
we cut out the region around S1 and continue the high-resolution simulation (Section \ref{subsec:high_reso_simulation}).}
Figure \ref{fig:SIGO_evolution_zoomin} shows the further evolution of
S1 from $z=25$ to $z = 21.4$, when it becomes Jeans unstable. The color map shows the gas number density around S1, and the color contours indicate DM mass overdensity. The distance between S1
and its closest DM halo is 0.9 kpc, which is four times larger than the halo's virial radius of 0.24 kpc.
%The distance between S1 and its closest DM halo and the virial radius the DM halo is almost the same as 0.11 physical kpc, and the baryon fraction in the Jeans radius of S1 is over 0.6. These values also indicate that S1 is not hosted by any surrounding DM halos.
%

Figure \ref{fig:zoomin_t_rho} shows the physical properties of S1 at $z=21.4$. \red{The top panel shows the gas-phase distribution in a density-temperature plane.}
%The temperature reaches $\sim 10^4 \, \mathrm{K}$, which is higher than
%the typical temperature of gas clouds $10^3 \, \mathrm{K}$
%in simulations without SV  \citep{Yoshida:2003, Yoshida:2006}.
%Because of SV, it takes time for the gas to contract and produce \ce{H2}. Thus, until \ce{H2} production has started, the temperature keeps rising due to adiabatic compression.}
It clearly indicates that \ce{H2} cooling is effective from $1~\mathrm{cm}^{-3}$, which enables S1 to condense and to collapse gravitationally;
the central gas density reaches $\sim 100~{\rm cm}^{-3}$ and the temperature is close to 200 K. In order to examine the gravitational instability of S1 quantitatively, we calculate the ratio of the enclosed gas mass to the Jeans mass (bottom panel). At radius $r = 50-100$
physical pc, the ratio is slightly above unity, suggesting that the cloud is Jeans unstable. We have followed further evolution of S1 until its density exceeds $10^{5}\, \mathrm{cm^{-3}}$ at $z = 20.0$. Runaway collapse is triggered shortly after S1 becomes Jeans unstable.

We also calculate the ratio of contraction time to free-fall time for S1. During the evolution shown in Figure \ref{fig:SIGO_evolution_zoomin}, the contraction time ($\rho/\dot{\rho}$) remains roughly the same as the free-fall time. The excess heat generated by the slow contraction of S1 is
radiated away by \ce{H2} cooling. Because the molecular fraction is reaching $f_{\rm H2} \sim 10^{-3}$ in S1, \ce{H2} cooling is effective, and S1 also contracts on the cooling time scale.

At the onset of gravitational collapse, the corresponding Jeans mass is $M_\mathrm{J, S1}\sim 10^5 M_\odot$, which is about $100$ times larger than that of a typical primordial gas cloud hosted by a DM mini halo. This is because the density when S1 becomes Jeans unstable is low ($\sim 100~{\rm cm}^{-3}$) owing to slow contraction, the time scale of which roughly follows the \ce{H2} cooling time.
It is interesting that the large Jeans mass is consistent with the conclusion of \citet{Peebles:1968}, who studied
the formation of primordial star clusters that are not hosted by dark halos.

\section{Discussion}\label{sec:discussion}
We have studied the formation and evolution of a cosmological SIGO.
We have shown, for the first time, that a SIGO cools and condenses by \ce{H2} cooling
to a very high density, when it becomes unstable to gravitational runaway collapse.
The massive SIGO is expected to form primordial stars.

The nature of SIGOs in cosmological simulations has been investigated in previous studies \citep[e.g.,][]{Popa:2016,Chiou:2018,Chiou:2019,Chiou:2021,Lake:2021}.
%(e.g., \citet{Popa:2016}, \citet{Chiou:2018}
%and \citet{Lake:2021}).
%They did not include \ce{H2} chemistry in their simulation, this the detected SIGOs did not reach Jeans instability.
In a recent study by \citet{Schauer:2021}, the formation and chemothermal evolution of SIGOs are studied in detail.
%However, they do not follow the collapse until the gas density reaches $\sim 100~{\rm cm}^{-3}$.
They conclude that \ce{H2} cooling alone is not sufficient for the gas clumps to condense down to low-enough temperatures for gravitational collapse.

In our simulations, \red{we have found} many SIGOs that end up with being hosted by nearby DM halos and also
SIGOs that do not condense by their own gravity and radiative cooling.
%First of all, in our simulation setup, we adopted an increased amplitude of density fluctuations by setting $\sigma_8 = 1.7$\citep{Chiou:2021} while \citet{Schauer:2021} used $\sigma_8 = 0.8159$ \citep{Plank:2016}. This makes our simulations more prone to gas contraction at higher redshifts. As a result, we are able to obtain more SIGO samples.
We have run high-resolution simulations for several SIGOs and have found that some SIGOs eventually cool down to 200\,K by \ce{H2} cooling, reaching the gas density of $\sim 100~{\rm cm}^{-3}$. We have found a SIGO (S1) that collapses to very high densities while being outside of DM halo(s).

 In our future study,  we will follow further evolution of the SIGO, which may even fragment and form a star cluster. Following the protostellar evolution of the member stars will reveal the properties of the first star cluster. Because the SIGO we have studied here is DM deficient, it is a promising candidate progenitor of a globular cluster.

Further studies involving the statistics of star-forming SIGOs and their typical properties such as mass, baryon fraction, etc., will clarify the relationship between SIGOs and globular clusters.
% It is also important to study the statistics of star-forming SIGOs and their typical properties such as mass, baryon fraction etc, in order to identify the relationship between SIGOs and globular clusters.

\section*{acknowledgments}
Y.N. would like to thank Shingo Hirano for fruitful discussions.
Numerical computations were carried out on Cray XC50 and PC cluster at Center for Computational Astrophysics, National Astronomical Observatory of Japan. N.Y. acknowledges financial support from JST AIP Acceleration Research JP20317829. S.N., W.L., and Y.S.C thank the support of NASA grant No. 80NSSC20K0500. S.N. thanks Howard and Astrid Preston for their generous support.

\bibliography{SIGOwithH2}{}
\bibliographystyle{aasjournal}

\end{document}